\documentclass[aps,pra,twocolumn,showpacs,superscriptaddress]{revtex4-1}

\usepackage{graphicx}
\usepackage{amsfonts}
\usepackage{amsmath}
\usepackage{amssymb}
\usepackage{url}
\usepackage{color}

\usepackage{bm}

\definecolor{dred}{rgb}{.8,0.2,.2}
\definecolor{ddred}{rgb}{.8,0.5,.5}
\definecolor{dblue}{rgb}{.2,0.2,.8}

\newcommand{\ket}[1]{\mbox{$|#1\rangle$}}
\newcommand{\ketbra}[2]{\mbox{$|#1\rangle\langle #2|$}}

\newcommand{\tr}{\mbox{tr}}					

\newcommand{\be}{\begin{equation}}
\newcommand{\ee}{\end{equation}}
\newcommand{\bea}{\begin{eqnarray}}
\newcommand{\eea}{\end{eqnarray}}

\begin{document}

\title{Quantum correlations in a noisy neutron interferometer}
\author{Christopher J. Wood}
\email{christopher.j.wood@uwaterloo.ca}
\affiliation{Institute for Quantum Computing, University of Waterloo, Waterloo, Ontario N2L 3G1, Canada}
\affiliation{Department of Physics and Astronomy, University of Waterloo, Waterloo, Ontario N2L 3G1, Canada}

\author{David G. Cory}
\affiliation{Institute for Quantum Computing, University of Waterloo, Waterloo, Ontario N2L 3G1, Canada}
\affiliation{Department of Chemistry, University of Waterloo, Waterloo, Ontario N2L 3G1, Canada}
\affiliation{Perimeter Institute for Theoretical Physics, Waterloo, Ontario N2L 2Y5, Canada}
\author{Mohamed O. Abutaleb}
\affiliation{Massachusetts Institute of Technology, Cambridge, Massachusetts 02139, USA }
\author{Michael G. Huber}
\affiliation{National Institute of Standards and Technology, Gaithersburg, Maryland 20899, USA}
\author{Muhammad Arif}
\affiliation{National Institute of Standards and Technology, Gaithersburg, Maryland 20899, USA}
\author{Dmitry A. Pushin}
\affiliation{Institute for Quantum Computing, University of Waterloo, Waterloo, Ontario N2L 3G1, Canada}
\affiliation{Department of Physics and Astronomy, University of Waterloo, Waterloo, Ontario N2L 3G1, Canada}
 
\date{\today}

\begin{abstract}

We investigate quantum coherences in the presence of noise by entangling the spin and path degrees of freedom of the output neutron beam from a noisy three-blade perfect crystal neutron interferometer. We find that in the presence of dephasing noise on the path degree of freedom the entanglement of the output state reduces to zero, however the quantum discord remains non-zero for all noise values. Hence even in the presence of strong phase noise non-classical correlations persist between the spin and path of the neutron beam. This indicates that measurements performed on the spin of the neutron beam will induce a disturbance on the path state. We calculate the effect of the spin measurement by observing the changes in the observed contrast of the interferometer for an output beam post-selected on a given spin state. In doing so we demonstrate that these measurements allow us to implement a quantum eraser, and a which-way measurement of the path taken by the neutron through the interferometer. While strong phase noise removes the quantum eraser, the spin-filtered which-way measurement is robust to phase noise. We experimentally demonstrate this disturbance by comparing the contrasts of the output beam with and without spin measurements of three neutron interferometers with varying noise strengths. This demonstrates that even in the presence of noise that suppresses path coherence and spin-path entanglement, a neutron interferometer still exhibits uniquely quantum behaviour.


\end{abstract}

\maketitle

\section{Introduction}
\label{sec:intro}

A unique property of quantum theory is that when two or more quantum systems are allowed to interact they may exhibit correlations that cannot be explained classically. In the field of quantum information science protocols harnessing these correlations can exceed classical efficiencies for certain metrology applications and information processing tasks \cite{Nielsen2000}. One of the most studied classes of correlated quantum states are entangled states as they enable extremely non-classical quantum effects such as quantum teleportation \cite{Horodecki2009}. A maximally-entangled quantum state of a bipartite quantum system allows for a projective measurement of one subsystem to completely determine the outcome of the corresponding projective measurements on the other. The class of states of interest to quantum computation however is broader then purely entangled quantum states, as certain non-entangled quantum states may still posses correlations that cannot be accounted for classically. In such cases measurement on one subsystem, while not determining the state of another, may still cause a disturbance to the state of the other. 

Classifying the quantum nature of correlations beyond entanglement has received much interest, with many discussions focused on quantum discord (QD) and related measures \cite{Lang2011,Celeri2011,Modi2011}. Quantum discord was proposed by Ollivier and Zurek \cite{Ollivier2002}, and Henderson and Vedral \cite{Henderson2001} to characterize quantum correlations in a bipartite system. In effect, one may interpret QD as a measure of the minimum disturbance that measurement of one subsystem of a bipartite quantum system can induce on the measurement outcomes of the other. Such classifications are of interest since certain quantum algorithms, such as DQC1, do not require entanglement to exceed classical efficiencies\cite{Knill1998}. It has been shown that for the DQC1 algorithm QD is present in the output state of the computation even when entanglement is not, and hence it was suggested that QD may provide a better figure of merit of evaluating quantum resources \cite{Datta2008}. Here we investigate the quantum nature of correlations of single neutrons in a neutron interferometer (NI).

Neutron interferometry has been used for precise tests of quantum mechanical phenomena such as coherent spinor rotation \cite{Rauch1975} and superposition \cite{Summhammer1983}, gravitationally induced quantum interference \cite{Colella1975}, the Aharonov-Casher effect \cite{Cimmino1989}, violation of a Bell-like inequality\cite{Hasegawa2003}, generation of a single neutron entangled state\cite{Hasegawa2007}, quantum contextually \cite{Bartosik2009}, and the realization of a Decoherence-Free subspace \cite{Pushin2011}. In our case a NI provides a clean system for considering quantum correlations in a bipartite quantum systems as we are able to coherently control the spin and path-momentum degrees of freedom of a neutron beam, and manipulate the correlations between them. In addition, due to the high efficiency of single neutron detectors and the low intensity of neutrons entering the interferometer, we are able to gather statistics from performing true projective measurements on single quantum systems. In the present article we investigate the correlations between the spin and path degrees of freedom of the output beam from a noisy NI by observing changes in the output beam intensity as a result of a post-selected projective measurement on the neutron spin. 

\section{Quantum Correlations}		\label{sec:qcorr}

\subsection{Quantum Discord}		\label{sec:discord}

Quantum discord is a non-symmetrical quantity defined by the difference between quantum generalizations of two classically equivalent expressions for \emph{mutual information}. Let $\rho_{AB}$ be a bipartite density matrix over two quantum systems $A$ and $B$. One expression for the mutual information of $\rho_{AB}$ is given by
\begin{equation}
I(A:B) = S(\rho_{A}) + S(\rho_{B})- S(\rho_{AB})
\label{eqn:mutual1}
\end{equation} 
where $S(\rho) = -\tr (\rho \log \rho)$ is the von-Neumann entropy of the density matrix $\rho$, $\rho_{A}=tr_B (\rho_{AB})$ is the reduced density matrix on subsystem $A$ taken by performing the partial trace over system $B$, and similarly $\rho_B=tr_A (\rho_{AB})$. An alternative expression for mutual information is formed by considering a quantum generalization of conditional entropy which accounts for possible measurement induced disturbances. Consider performing a measurement on subsystem $B$, this is most generally described by a positive operator valued measure (POVM) $E$ consisting of a set of measurement operators $\{E_b\}$ satisfying $E_b\ge 0$, $\sum_b E_b = \bm {1}$ \cite{Nielsen2000}. Measurement outcome $b$ will occur with probability $p_b = \tr(E_b \rho_{AB})$, and the post-measurement state of subsystem $A$, conditioned on outcome $b$, is given by 
\be
\rho_{A|b}=\frac{1}{p_b}\tr_B (E_b \rho_{AB}).
\ee
We may define a generalization of conditional entropy for a given POVM $E$ as 
\be
S(\rho_{A|E})= \sum_b p_b S(\rho_{A|b}).
\ee
This gives us an alternative expression for mutual information by maximizing over all possible POVMs:
\be
J(A|B) = \max_{E} \left[ S(\rho_A) -S(\rho_{A|E}) \right]
\label{eqn:mutual2}
\ee

Quantum discord is defined to be the difference between expressions \eqref{eqn:mutual1} and \eqref{eqn:mutual2}:
\begin{eqnarray}
D(A|B) &=& I(A:B) - J(A|B) \nonumber\\
&=& \min_{E}\Big[	S(\rho_{A|E}) + S(\rho_{B}) - S(\rho_{AB})		\Big]
\label{eqn:discord}
\end{eqnarray}
Similarly one may define the quantum discord $D(B|A)$ where one optimizes over POVMs on subsystem $B$. 
In general to compute the quantum discord of a state one must minimize Eq.~\eqref{eqn:discord} over all extremal rank-one POVMs, however it has been shown that for rank-two states orthogonal projective valued measurements (PVMs) are optimal \cite{Galve2011}.

\subsection{Entanglement of Formation}		\label{sec:eof}

There are numerous measures for quantifying entanglement in a quantum state (for a review of entanglement see \cite{Horodecki2009}). In our case a convenient measure for a two-qubit mixed-state is the \emph{entanglement of formation} (EOF)~\cite{Wootters1998} which is given by
\be
EOF(\rho_{AB})	=	h\left(\frac{1+\sqrt{1-\mathcal{C}(\rho_{AB})^2}}{2}\right)	
\ee
where $h(x) = -x \log x -(1-x)\log(1-x)$, and $\mathcal{C}(\rho_{AB})$ is the \emph{concurrence} of a bipartite state $\rho_{AB}$:
\be
\mathcal{C}(\rho_{AB}) = \max\{0, \lambda_1 - \lambda_2 -\lambda_3-\lambda_4\}
\ee
where $\lambda_j$ are the eigenvalues of the Hermitian matrix
\be
\sqrt{\sqrt{\rho_{AB}}(Y\otimes Y)\rho_{AB}^* (Y\otimes Y)\sqrt{\rho_{AB}}}
\ee
sorted such that $\lambda_1 \ge \lambda_2 \ge \lambda_3 \ge  \lambda_4$, where $*$ denotes complex conjugation, and $Y$ is the Pauli-$Y$ matrix.

\subsection{Quantum Correlations in a Neutron Interferometer}		\label{sec:qcorr-ni}

We now consider quantum correlations in the output state of a three-blade NI and will follow with the mathematical model of the NI used to derive them Section~\ref{sec:ni-model}. In our configuration systems $A$ and $B$ correspond to the path and spin degrees of freedom of a neutron respectively, which each may be modelled as a two-level quantum system (qubit). By performing a controlled spin-rotation of angle $0\le\alpha\le2\pi$ in one of the paths of the NI we may introduce entanglement between the spin and path subsystems of an initially spin-polarized neutron beam. In a realistic NI there are noise sources which introduce decoherence and reduce the effectiveness of this entangling operation. In the present paper we consider the decoherence due to surface defects of the NI blades. This noise source introduces a random phase between the two interferometer paths which degrades the coherence of the path subsystem $A$. 

\begin{figure*}[htbp]
\centering
\includegraphics[width=0.75\textwidth]{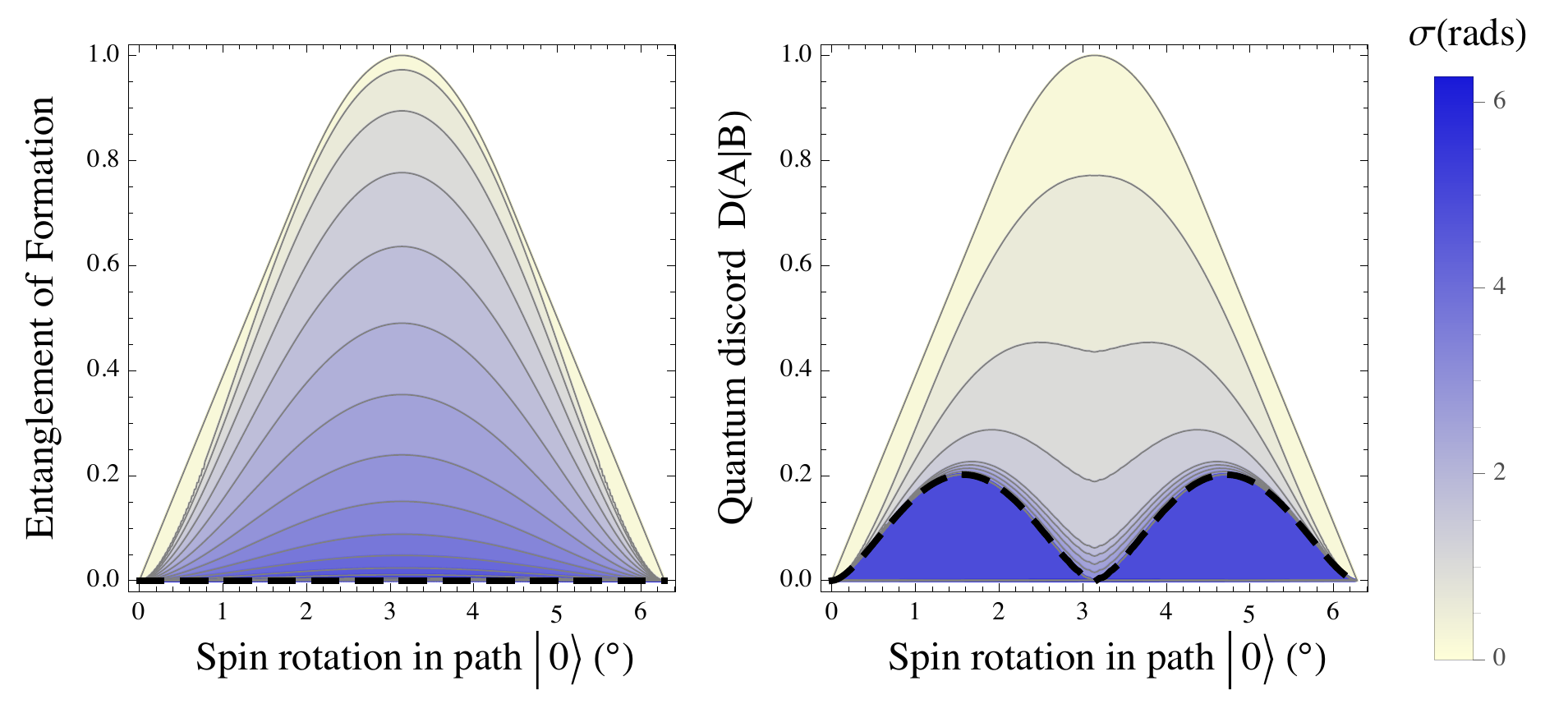}
\caption{(Colour online) Entanglement of formation (left) and quantum discord $D(A|B)$ (right) between the spin and path degrees of freedom of neutrons exiting a three-blade NI as a function of the spin rotation angle of neutrons in the $\ket{0}$ interferometer path, and noise strength $\sigma$. The NI schematic is described in Fig. \ref{fig:ni-setup}, and the noise model considered introduces a normally distributed random phase, with mean 0 and standard deviation $\sigma$, between the NI paths. The dashed line corresponds to the maximum noise case of a uniform distribution of angles. While the entanglement approaches zero for all spin rotation angles as the noise strength increases, the quantum discord remain non-zero.}
\label{fig:conc-qd}
\end{figure*}

Since the neutrons exiting the NI may be described by a mixed state of a 2 qubit quantum system, we use EOF as a measure of the entanglement in the output state. Further, since the quantum state of the neutrons is rank-two we need only perform the minimization in Eq.~\eqref{eqn:discord} over PVMs on the spin subsystem to calculate the quantum discord. We find that the EOF between the spin and path systems goes to zero asymptotically as the strength of the random phase noise increases, while the QD remains non-zero for all values of the spin rotation except $n\pi$ for integer values of $n$. This is illustrated in Fig. \ref{fig:conc-qd}. Even though there is no entanglement between the spin and path of the neutrons in the case of strong phase noise, the non-zero quantum discord $D(A|B)$ indicates the presence of non-classical correlations. This signifies that measurements performed on the neutron spin will induce a disturbance on the path state of the output neutron beam.

The observed entanglement evolution under an increase in the strength of the phase noise can be classed as \emph{approaching} \cite{Zhou2012}, in contrast to entanglement sudden death \cite{Yu2009}. QD has been shown to be robust to sudden death and instead asymptotically vanishes in bipartite systems subject to Markovian evolution \cite{Werlang2009,Ferraro2010}, however in our case QD remains asymptotically non-zero for most spin rotation values. Similar effects of the vanishing of entanglement but non-vanishing QD have been previously found in the theoretical analysis of the evolution of coupled quantum dots under decoherence \cite{Fanchini2010}. Certain initially correlated two-atom states have also been shown to have a non-vanishing quantum discord when coupled to a common dissipative cavity \cite{ZhangYJ2011}.

\section{Theoretical Model}		\label{sec:ni-model}

\begin{figure}[htbp]
\centering
\includegraphics[width=0.48\textwidth]{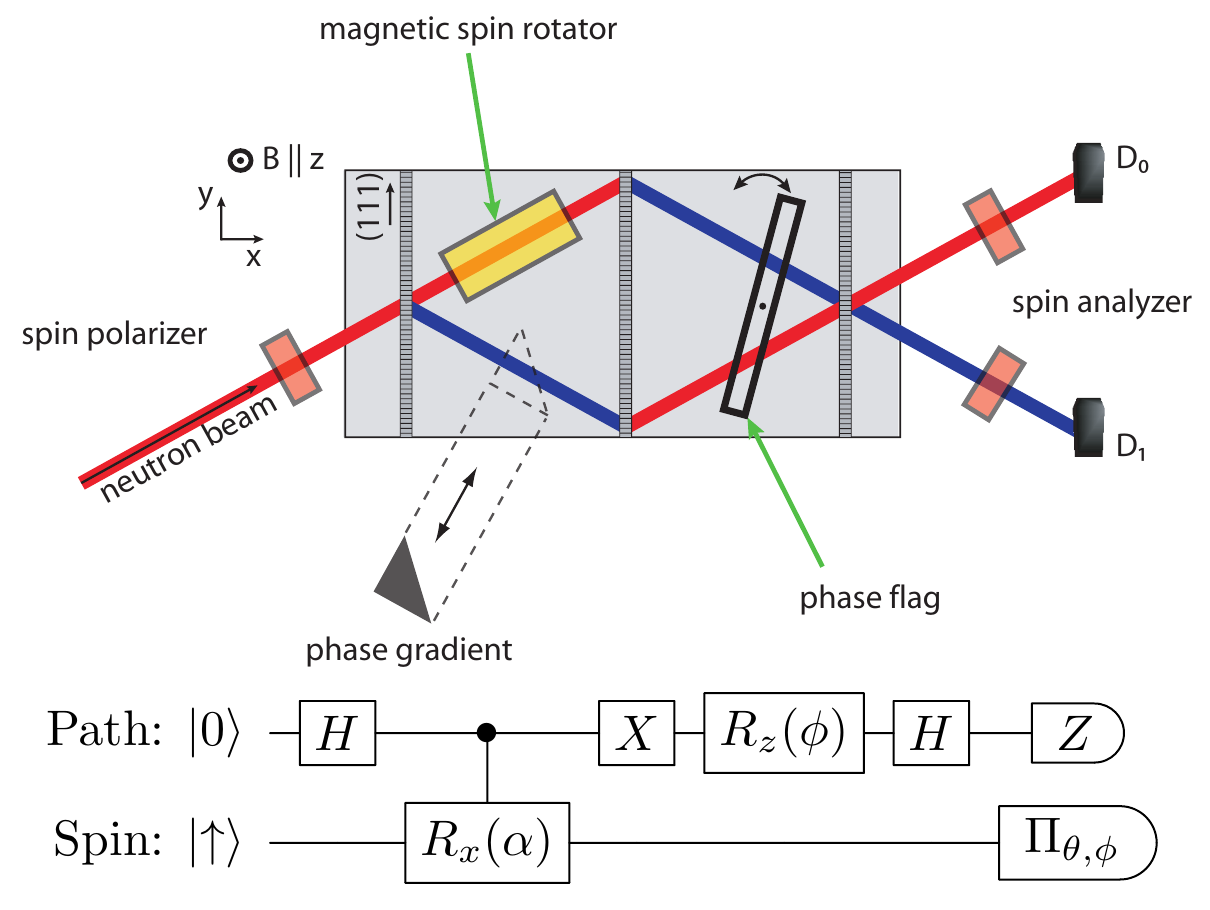}
\caption{(Color online) Experimental setup for the three blade neutron interferometer (top) and the corresponding quantum circuit for the ideal model (bottom). The red (blue) paths in NI schematic are defined as the $\ket{0}$($\ket{1}$) path states, $H$ is a Hadamard gate, $R_x(\alpha)$ is a rotation of the neutron spin in the $\ket{0}$ path of $\alpha$ radians about $x$-axis, $X$ is a bit-flip, $R_z(\phi)$ is a relative phase shift of $\phi$ radians between the beam paths, $\Pi$ is a projective measurement performed on the spin-state (spin-analyser) in the basis $\cos(\theta)\ket{\uparrow}\pm e^{i\phi}\sin(\theta)\ket{\downarrow}$, and $Z$ is a projective measurement of the path intensities in the $\ket{0},\ket{1}$ basis.}
\label{fig:ni-setup}
\end{figure}

We will now briefly describe the mathematical model used to describe the neutron interferometer. The most common geometry for a NI is a three-blade system machined from a perfect single crystal of silicon. This functions as a Mach--Zender interferometer on the longitudinal momentum of the neutron beam. We refer to this degree of freedom of the neutron beam as the \emph{path} system. The neutron path can be viewed as a two level system which we may couple to the neutron spin to form a bipartite quantum system. In this context we may view the interferometer crystal as a quantum circuit acting as illustrated in Fig. \ref{fig:ni-setup}. We define the basis for the path to be the computational basis where $\ket{0}$ and $\ket{1}$ correspond to the red and blue beam paths in Fig~\ref{fig:ni-setup} respectively. For the spin-system we work in the spin-up, spin-down eigenbasis $\ket{\uparrow}, \ket{\downarrow}$ with respect to a static field in the $z$-direction. 

The first (and third) NI blades act as Hadamard ($H$) gates on the neutron path by coherently splitting (and recombining) the neutron beam into two paths via Bragg scattering in the Laue geometry \cite{Sears1989}. The second NI blade deflects the beam by swapping the path-momentum directions, which we model as a bit-flip ($X$) gate. In our defined bases these are given by: 
\bea
H &=& \frac{1}{\sqrt{2}}\left(\ketbra{0}{0}+\ketbra{0}{1}+\ketbra{1}{0}-\ketbra{1}{1}\right) \\
X &=& \ketbra{0}{1}+\ketbra{1}{0} 	
\eea
In practice the intensity of the output neutron beam is reduced due to neutrons escaping the NI at the second blade, however we account for this in our description of the output beam by post-selecting on the neutrons which remain in the interferometer. 

Between the first and second NI blades we couple the spin and path degrees of freedom by selectively rotating the neutron spin in the $\ket{0}$ path by an angle $\alpha$. This acts a controlled-$X$ rotation ($R_x(\alpha)$), with the spin and path as the target and control respectively:
\bea
\mbox{C-}R_x(\alpha) &=&	
	\ketbra{0}{0}
	\otimes
	R_x(\alpha)
	+
	\ketbra{1}{1}
	\otimes \bm{1}_s
\\	
R_x(\alpha) &=& 	
		\exp\left[i\,\frac{\alpha}{2}\,\left(
		\ketbra{\uparrow}{\downarrow}+
		\ketbra{\uparrow}{\downarrow}\right)\right] 
		\\
\bm{1}_s &=& \ketbra{\uparrow}{\uparrow}+\ketbra{\downarrow}{\downarrow}
\eea

We measure the intensities of the output beams using two $^3$He integrating detectors called $D_0$ and $D_1$, corresponding to projective measurements of the states $\ket{0}$ and $\ket{1}$ respectively. This performs a $Z$-basis measurement on the neutron path subsystem. By including spin-filters which selectively transmit neutrons with a preferred spin we may also perform post-selected spin measurements. This allows us to perform joint measurements on the spin and path of the neutron beam.

In a typical NI experiment a relative phase of $\phi$ is induced between the two paths by a phase flag between the second and third blades which effectively implements the Z-rotation gate:
\be
R_z(\phi) = 	
e^{-i\phi/2}  \ketbra{0}{0} + e^{i\phi/2}\ketbra{1}{1}.	
\ee
The relative phase $\phi$ parameterizes the measured beam intensity by controlling the interference between the two beam paths recombined at the third blade. 

Ideally the input beam is in the spin-up polarized state $\psi_{in}= \ket{0}\otimes\ket{\uparrow}$  with respect to a uniform magnetic field in the $z$-direction. In practice however one is not able to perfectly polarize the input neutron beam and in general we describe the input beam by the state
\begin{equation}
\rho_{in}(\epsilon) = \ketbra{0}{0}\otimes\left(\frac{1+\epsilon}{2} \ketbra{\uparrow}{\uparrow}+\frac{1-\epsilon}{2}\ketbra{\downarrow}{\downarrow}\right)
\label{eqn:rhospin}
\end{equation}
where $-1\le\epsilon\le1$ parameterizes the spin-polarization of the neutron beam.

\subsection{Output Intensities}
\label{sec:output-int}

In an ideal neutron interferometer interference effects are observed in the measured output intensity at each detector. The ideal output intensity is a function of the relative phase between interferometer paths and the angle of spin rotation in the $\ket{0}$ path. If no measurement is performed on the neutron spin subsystem, the ideal detector probabilities in the absence of noise are given by
\bea
D_{0,\scriptsize\mbox{Ideal}}(\phi,\alpha) &=&		\frac{1}{2}\left[1+\cos\left(\frac{\alpha}{2}\right)\cos(\phi)\right]	
\nonumber \\
D_{1,\scriptsize\mbox{Ideal}}(\phi,\alpha) &=&		\frac{1}{2}\left[1-\cos\left(\frac{\alpha}{2}\right)\cos(\phi)\right].
\label{eqn:detector}
\eea
which is independent of the spin-polarization of the neutron beam. In a real NI the blades do not generally have equal transmission and reflection coefficients and hence are not true 50-50 beam splitters. This doesn't effect the interference effects at detector $D_0$ though since both interferometer paths to this detector have the same number of transmissions and reflections.

In practice neutron interferometers cannot be machined perfectly and surface imperfections in the crystal blades lead to a distribution of phases over the cross-sectional area of the neutron beam. This results in reduced contrast of the beam intensity when averaged over the beam distribution. To include the effect of phase noise in our model we consider the output intensities with a phase shift $\phi+\phi_r$ where $\phi_r$ is an additional random phase shift introduced between paths by the NI blades. This random phase is assumed to be normally distributed with mean 0 and variance $\sigma$. By averaging over the distribution of $\phi_r$, we may obtain the average detector intensities:
\bea
D_{0}(\phi,\alpha,\sigma) 
	&=&	\int_{-\infty}^{\infty}d\phi_r\, D_{0,\scriptsize\mbox{Ideal}}(\phi+\phi_r,\alpha) \, 
		\frac{\exp\left(-\frac{\phi_r^2}{2\sigma^2}\right)}{\sqrt{2\pi\sigma^2}}
\nonumber\\
D_{0}(\phi,\alpha,\sigma) 	&=& \frac{1}{2}\left[1+e^{-\sigma^2/2}\cos\left(\frac{\alpha}{2}\right)\cos(\phi)\right]	
	\label{eqn:d0-noise}
\eea

\subsubsection{Output intensity with spin-filtering}

We now consider the detector intensities and contrast curves when we include a spin-filter to perform a post-selected spin measurement on the output neutron beam before detector $D_0$. The spin-filter implements a post-selected projective measurement of the pure state
\be
\ket{S(\theta,\varphi)} = \cos\left(\frac{\theta}{2}\right)\ket{\uparrow}
				+e^{i\varphi}\sin\left(\frac{\theta}{2}\right)\ket{\downarrow}
\label{eq:filterstate}
\ee
where $\theta, \phi$ are the spherical coordinates parameterizing the state on the Bloch Sphere. In practice the spin-filter acts by absorbing neutrons in the orthogonal spin state before they reach the detector. After post-selection the detector intensity is proportional to
\bea
&&D_{0,S(\theta,\varphi)}(\phi,\alpha,\sigma)
= \frac{1}{4} 
	\bigg(
	1+\epsilon\cos^2\left(\alpha/2\right) \cos(\theta)
\nonumber\\
&&
	+\frac{\epsilon}{2} \sin(\alpha) \sin(\theta) \sin(\varphi)
\nonumber\\
&&
	+e^{-\frac{s^2}{2}} \cos(\alpha/2) (1+\epsilon\cos(\theta)) \cos(\phi)
\nonumber\\
&&
	-e^{-\frac{s^2}{2}}\sin(\alpha/2) \sin(\theta)\cos(\phi)\big[\sin(\phi) -\epsilon\sin(\varphi)\big]
\bigg)
\eea

We explicitly consider two cases, spin-filtering in the same axes as the quantizing magnetic field ($Z$-filter), and spin-filtering in an orthogonal basis ($X$-filter). These are given by
\be
\begin{array}{lll}
Z: 	&	\ket{\uparrow} = \ket{S(0,0)}		
	&  	\ket{\downarrow} = \ket{S(\pi,0)} 	\\
X: 	&	\ket{\uparrow x} = \ket{S(\pi/2,0)} 
	&	\ket{\downarrow x} = \ket{S(3\pi/2,0)}
\end{array}
\ee
in terms of the $(\theta,\phi)$ parameterization in \eqref{eq:filterstate}.

In these case of the $Z$-filter the observed intensities at detector $D_0$ are proportional to
\bea
D_{0,\uparrow z}(\phi,\alpha,\sigma,\epsilon)
	&=&  \left(\frac{1+\epsilon}{2}\right)\, D_{0}(\phi,\alpha,\sigma)
		\nonumber\\&&
		+\frac{\epsilon}{8}\left[\cos(\alpha)-1\right]	
		\label{eqn:d0-z}	\label{eqn:d0upz}\\
D_{0,\downarrow z} (\phi,\alpha,\sigma,\epsilon)
	&=&  \left(\frac{1-\epsilon}{2}\right)\, D_{0}(\phi,\alpha,\sigma)
		\nonumber\\&&
		-\frac{\epsilon}{8}\left[\cos(\alpha)-1\right]	
		\label{eqn:d0-z}	\label{eqn:d0downz}
\eea
for spin-up and spin-down filtering in the $z$-direction respectively. Note that in this case the normalization condition for the output probabilities is that $D_{0,\downarrow} + D_{0,\uparrow}+D_{1,\downarrow} + D_{1,\uparrow}=1$.

In these cases of the $X$-filter the observed intensities at detector $D_0$ are proportional to
\bea
D_{0,\uparrow x}(\phi,\alpha,\sigma)
	&=&	\frac{1}{4}\left[1+ e^{-\sigma^2/2}\cos\left(\frac{\alpha}{2}+\phi\right)\right]
		\label{eqn:d0-xu}	\\
D_{0,\downarrow x}(\phi,\alpha,\sigma)
	&=&	\frac{1}{4}\left[1+ e^{-\sigma^2/2}\cos\left(\frac{\alpha}{2}-\phi\right)\right]
		\label{eqn:d0-xd}
\eea
for spin-up and spin-down filtering in the $x$-direction respectively. 
We see here that the the $Z$-filter adds an additional term to the unfiltered contrast, while the $X$-filtering combines the parameters $\phi$ and $\alpha$ into a single argument of a cosine function. Further, in the weak noise case ($\sigma\approx 0$) both these expressions are observably different from the non-spin-filtered case in Eq.~\eqref{eqn:d0-noise}. However in the case of strong noise, the non-spin-filtered and $X$-filtered intensities approach constant values. Only the $Z$-filtered intensities are observably different to the non-spin-filtered intensity, and depend on the initial spin polarization $\epsilon$, and the controlled spin-rotation angle $\alpha$. We discuss the implications of these results in Section~\ref{sec:interp}, but first we introduce a measure of coherence in interferometer experiments called \emph{contrast}.

\subsection{Contrast}
\label{sec:contrast}

The intensity curves for each detector as a function of the relative phase $\phi$ between interferometer paths are referred to as \emph{contrast curves}. They are analogous to the interference pattern produced by a double slit interference experiment. The difference between the maximum and minimum intensity of the $D_0$-detector as a function of a phase-flag rotation $\phi$ is called the \emph{contrast} of the NI and is defined as
\begin{equation}
C_{P} = \frac{ \max_{\phi}[D_0(\phi)]-\min_{\phi}[D_0(\phi)]}{\max_{\phi}[D_0(\phi)]+\min_{\phi}[D_0(\phi)]}.
\label{eqn:contrast}
\end{equation}
The contrast may take values $0\le C_{P}\le1$ and is a measure of the strength of quantum coherence between the paths.

We also consider an alternative contrast expression where our parameter of variation in detector intensity is the angle of spin rotation $\alpha$ rather than the phase rotation $\phi$ as in Eq.~\eqref{eqn:contrast}. We define the \emph{spin-contrast} to be
given by
\be
C_{S} = \frac{ \max_{\alpha}[D_0(\alpha)]-\min_{\alpha}[D_0(\alpha)]}{\max_{\alpha}[D_0(\alpha)]+\min_{\alpha}[D_0(\alpha)]}.
\label{eqn:spin-contrast}
\ee
We will refer to the standard contrast as the \emph{path-contrast} to distinguish it from the spin-contrast. 

\subsubsection{Contrast without spin-filtering}

Using the observed detector probability in Eq.~\eqref{eqn:d0-noise} we may calculate that the path and spin contrasts of the noisy three blade NI:
\bea
C_{P}(\alpha,\sigma) 	&=& e^{-\sigma^2/2}\,
						\left|\cos\left(\frac{\alpha}{2}\right)\right|
						\label{eqn:path-contrast}\\
C_{S}(\phi,\sigma) 	&=& e^{-\sigma^2/2}\,
						\left|\cos\left(\phi\right)\right|
						\label{eqn:spin-contrast}
\eea
We see here that the average contrast and spin-contrast expressions are equivalent but with the roles of $\alpha$ and $\phi$ interchanged ($C_{S}(\phi,\sigma)=C_{P}(2\phi,\sigma)$), and depend on the noise strength, and the phase of the parameter that is not optimized over for the contrast ($\alpha$ for path contrast and $\phi$ for spin-contrast).

\subsubsection{Contrast with spin-filtering}

We now consider the theoretical path and spin-contrasts of the output beam after spin-filtering. When we post-select on the spin-up and spin-down states of the $X$-filter we obtain contrast values of
\bea
C_{P(\uparrow x)}(\sigma) &=& C_{S(\uparrow x)}(\sigma) = e^{-\sigma^2/2}	
\label{eqn:cp-x-up}
\\
C_{P(\downarrow x)}(\sigma) &=& C_{S(\downarrow x)}(\sigma) = e^{-\sigma^2/2}.
\label{eqn:cp-x-down}
\eea
We find that the spin and path contrasts are equivalent and depend only on the strength of the phase noise. In particular the contrast decreases to zero with the increase in noise strength.

For the $Z$-filtered intensities we obtain post-selected path-contrasts of
\bea
C_{P(\uparrow z)}(\alpha,\sigma,\epsilon)&=&\left|\frac{(1+\epsilon) e^{-\sigma^2/2}\cos\left(\frac{\alpha }{2}\right)}{1+\frac{\epsilon}{2}(1+\cos(\alpha))}\right|
	\label{eqn:cp-z-up}\\
C_{P(\downarrow z)}(\alpha,\sigma,\epsilon)&=&\left|\frac{(1-\epsilon) e^{-\sigma^2/2}\cos\left(\frac{\alpha }{2}\right)}{1-\frac{\epsilon}{2}(1+\cos(\alpha))}\right|.
	\label{eqn:cp-z-down}
\eea
which satisfy
\be
C_{P(\uparrow z)}(\alpha,\sigma,\epsilon)
\ge C_{P}(\alpha,\sigma)
\ge C_{P,\downarrow z}(\alpha,\sigma,\epsilon)
\ee
for $\epsilon\ge0$, with equality in the case of an zero spin-polarization ($\epsilon = 0$). In particular we see that $C_{path,\downarrow z}(\alpha,\sigma,1) = 0$.

The spin-contrasts for the $Z$-filtered intensities are more complicated as the values of $\alpha$ which obtain the minimum for the detector intensities are in general functions of $\phi, \epsilon$ and $\sigma$. For the spin-up Z-filter we have
\bea
C_{S(\uparrow z)}(\phi,\sigma,\epsilon)&=&
	\frac{\epsilon+(1+\epsilon)C_{S}(\phi,\sigma)+C_{S}(\phi,\sigma)^2}
	{2+\epsilon+(1+\epsilon)C_{S}(\phi,\sigma)-C_{S}(\phi,\sigma)^2}.
\nonumber\\	\label{eqn:cs-z-up}
\eea
For the spin-down Z-filter, in the range of $\frac13\le\epsilon\le 1$, we have
\bea
C_{S(\downarrow z)}(\phi,\sigma,\epsilon)&=&
	\frac{
	\epsilon-(1-\epsilon)C_{S}(\phi,\sigma)+\frac{(1-\epsilon)^2}{4\epsilon}C_{S}(\phi,\sigma)^2}
	{2-\epsilon+(1-\epsilon)C_{S}(\phi,\sigma)-\frac{(1-\epsilon)^2}{4\epsilon}C_{S}(\phi,\sigma)^2}.
	\nonumber\\
		\label{eqn:cs-z-down}
\eea

For the specific case of unpolarized neutrons ($\epsilon = 0$) we have 
\be
C_{S(\uparrow z)}(\phi,\sigma,0) 
	= C_{S(\downarrow z)}(\phi,\sigma,0)
	= C_{S}(\phi,\sigma)
\ee
and in the case of perfect polarization ($\epsilon=1$) we find that for spin-down Z-filtering we have perfect spin-contrast:
\be
C_{S(\downarrow z)}(\phi,\sigma,1) = 1
\ee

In the case of strong noise the Z-filtered spin-contrast expressions reduce to 
\bea
C_{S(\uparrow z)}(\phi,\infty,\epsilon)&=&
	\frac{\epsilon}{2+\epsilon}\\
C_{S(\downarrow z)}(\phi,\infty,\epsilon)&=&
	\frac{\epsilon}{2-\epsilon}.
\eea
and depend only on the initial polarization $\epsilon$ of the neutron beam. In practice strong noise amounts to $\sigma\ge2\pi$.

\section{Interpretation and Proposed Experiments}
\label{sec:interp}

We now discuss the significance of previously calculated path-constrast and spin-contrast values for the noisy 3-blade neutron interferometer. In the absence of spin-filtering, while both the path-contrast and spin-contrast of the ideal 3-blade NI go to zero as the noise strength $\sigma$ increases, as shown in Fig \ref{fig:conc-qd}, there is a non-zero quantum discord $D(A|B)$ between the spin and path subsystems. This implies that if we implement a measurement on the spin system the output intensities of the path system must be affected. By using a spin-filter we are able to post-select on an outcome arbitrary PVM on the spin neutron system, however to observe the influence of the spin-filter we are restricted by only being able to measure the path subsystem in the $\ket{0}, \ket{1}$ basis due to the inability to change the final blade of the NI. Hence when restricted to a single measurement basis this influence may not be observable for all spin post-selected states.

\subsection{No spin post-selection}

In the absence of spin-filtering we found that the path-contrast for the noisy neutron interferometer as given in Eq.~\eqref{eqn:path-contrast}, dependent only on the noise strength $\sigma$ and the angle of controlled spin-rotation $\alpha$. In the absence of noise, as we increase the angle of spin-rotation up to a $\alpha=\pi$ the measured contrast reduces to zero. At $\alpha=\pi$ the spin and path subsystems are maximally entangled, as shown by an EOF of 1 in Fig~\ref{fig:conc-qd}. By not measuring the spin subsystem we are partial trace over this subsystem which, in the case of a maximally entangled state, results in a maximally mixed reduced state of the path subsystem, and hence zero contrast. This may be interpreted as having performed a \emph{which-way} measurement of the path taken by the neutron through the interferometer. The neutrons passing the spin-filter are marked to have spin-down, while the neutrons which don't go through the arm with the spin-rotator will all have spin-up. By tuning $0<\alpha<\pi$ we may control the strength of this which-way marking of the neutrons. For $\alpha$ close to 0 it becomes a weak which-way marking of the path taken by the neutrons through the interferometer, and hence we still retain some contrast.

In the case of spin-contrast, as given in Eq.~\eqref{eqn:spin-contrast}, we have the same situation but with the roles of the rotation angle $\alpha$ and phase-flag $\phi$ reversed. In this case we are doing a spin-based magnetic interference experiment, and the relative phase between paths now performs the which-way marking of the neutron. In  both cases the presence of noise reduces the value of contrast, until it is approximately zero at $\sigma=2\pi$. This would suggest that the random phase noise destroys all relative phase information, and hence coherence, between the two paths in the interferometer. However due to the non-zero discord between the path and spin of the neutron we may attempt to recover some information by spin-measurements.

\subsubsection{$X$-filter spin post-selection}

In the case of $X$-filtering we found that both the path-contrast and spin-contrast when spin-filtering on the $\ket{\uparrow x}$ or $\ket{\downarrow x}$ spin-states dependend only on the strength $\sigma$ of the random-phase noise, as shown in Eq.\eqref{eqn:cp-x-up}. This is because the $X$-filter post-selection acts to combine the parameters $\alpha$ and $\phi$ into a single relative phase parameter $\phi+\alpha/2$ between the two NI paths which is observed at the detector. For path-contrast the spin rotation angle only shows up as a shift of the contrast curves without changing the actual contrast value. In effect the X-filter has erased the which-way marking of the neutrons in the NI due to the controlled spin-rotation angle. Similarly, for the spin-contrast the roles of $\phi$ and $\alpha$ are swapped with the spin-filter now erasing all effect of the phase-flag parameter on the output intensities. This is analogous to a \emph{quantum eraser} in optics~\cite{Scully1982}. By post-selecting on the neutron spin in the $x$-direction we have erased the which-way measurement caused by entanglement between the spin and path neutron subsystems. However, as the noise strength increases the spin-filtered spin and path contrasts both reduce to zero and become indistinguishable form the unfiltered contrast.

\subsubsection{$Z$-filter spin post-selection}

When implementing a $Z$-filter post-selection we calculated quite different values for the spin-contrast and path-contrast. In the case where we post-select on the $\ket{\uparrow}$ spin-state, the path-contrast in Eq.~\eqref{eqn:cp-z-up} is maximized for a perfectly polarized input ($\epsilon=1$), as we are in effect filtering out only the portion of neutrons rotated away from $\ket{\uparrow}$ by the spin-rotator. In this sense, much like the non-post-selected case, the angle of rotation controls the strength of the which-way measurement. If instead we filter on $\ket{\downarrow}$ as given by Eq.~\eqref{eqn:cp-z-down}, then we find the contrast is zero for $\epsilon = 1$. This is because in this case we are post-selecting only on the spins that were rotated, and hence we are performing a perfect which-may measurement of the path taken by the neutrons and cannot have any path based interference effects . If the incoming beam is not perfectly polarized our which-way measurement is effectively noisy and we will have a fraction of unrotated neutrons which still have spin-down polarization. In this case, as with the $\ket{\uparrow}$ filter, the angle of rotation $\alpha$ controls the relative strength of the which-way measurement.

For the spin-filtered spin-contrast we find that with perfect polarization the spin-filtered contrast with spin-down post selection is always 1. However with $\epsilon < 1$ the value of contrast will depend on the phase-flag $\phi$, which acts as the which-way marking. As with the unfiltered case it will be maximum for $\phi =0$, and zero at $\phi = \pi/2$.

As the noise strength increases, the dependence of $\phi$ is removed as we are decohering the relative phase information between paths. Hence the noise is erasing the which-way marking due to the phase-flag on the spin-contrast. In the strong noise case we find that the spin-contrast depends only on the initial polarization $\epsilon$. If $\epsilon=1$ we are filtering out all spins that are not rotated to $\ket{\downarrow}$, so our measured intensity is a function of the rotation angle. In the $\epsilon <1$ case, we are effectively introducing spin noise into the system as there will now be $(1-\epsilon)/2$ portion of neutrons with spin-down in the non-rotated path, thus reducing the spin-contrast. For the spin-up filter we have a similar situation, however the contrast is no longer unity for $\epsilon=1$ unless $\phi=0$ and $\sigma=0$. In this case we are filtering out the percentage of neutrons rotated to spin-down by the spin-rotator, rather than post-selecting on them. 

It has been suggested this setup might be used to demonstrate the so-called \emph{quantum cheshire cat} paradox~\cite{Aharonov2013}. This paradox is to weakly perform two measurements of the path a neutron takes through the interferometer simultaneously. One that is spin-based and determines that the neutron spin goes down one arm of the interferometer, and another that is not spin-based and determines that the neutron itself went down the other interferometer arm. By doing the which-way marking with a spin-filter we may measure which path the neutron spin went down by a spin based measurement. By varying $\alpha$ we may control the strength of this measurement. To complete the experiment would require implementing a second weak measurement simultaneously to suggest that the neutron itself was observed to go down a different path to its spin degree of freedom. This has been suggested to be implemented in a NI by using a partial absorber in the interferometer path without the spin rotator \cite{Denkmayr2013}.

\section{Experimental Demonstration}

We were able to experimentally demonstrate some of the theoretical results from Section \ref{sec:contrast}. However, temperature variations caused by the the method of implementing the spin-rotator resulted in a phase drift which increases the effective strength of the phase noise in our NI. Hence while we can calculate path-contrast in the absence of spin-rotation, for the spin-contrast experiments the spin-rotator acts to increase the apparent phase noise so that $e^{-\sigma^2/2} \approx 0$. With the increase in phase noise the spin-contrast with no spin-filtering, and with X-filtering is expected to be approximately zero. We may only observe the $Z$-filtered contrast which in the strong noise case only depends on the neutron polarization. This $Z$-filter post-selection demonstrates the disturbance of the path state of the neutrons by measurement of the neutron spin in the presence of strong noise, as indicated by the non-zero QD as shown in Fig.~\ref{fig:conc-qd}. 

\subsection{Experimental Setup}

We compared the contrast and spin-filtered contrasts after post-selecting on spin-down neutrons, quantized in a static magnetic field in the $z$-direction, for three NIs using the setup shown in Fig. \ref{fig:ni-setup}. The experiment was performed at the National Institute of Standards and Technology Center for Neutron Research's Neutron Optics and Interferometer Facility, located at Gaithersburg, Maryland~\cite{NistWeb}. This facility has an excellent vibration isolation and temperature stability thus allows for a good and long phase stability~\cite{Pushin2008}. 

Our neutron beam consisted of 0.271nm wavelength neutrons, and the incident neutron beam was polarized via a transmission mode supermirror polarizer~\cite{Abutaleb2012} giving an initial polarization of $93\%$ spin-up. The path-selective spin rotation was implemented using thin permalloy films~\cite{PynnFootnote} deposited on Si substrate~\cite{Pynn2005}. Spin-filters were implemented using either Heusler crystals or reflection-mode curved supermirrors. These were preceded by an adiabatic coil used to rotate the neutron spin so that spin-up neutrons were absorbed, and spin-down neutrons were transmitted. During this experimental work we have used two LLL type NI with different initial contrasts: ``good'' and ``bad'', which we refer to as $N_1$ and $N_2$ respectively. To compare spin-contrast with a very low contrast NI under the same environmental conditions we used the good NI and introduced a large destructive phase gradient by adding a 45 degree fused silica wedge in one interferometer path~\cite{Pushin2007}. We refer to the good NI with the wedge as $N_3$.

\begin{figure}[htbp]
\begin{center}
\includegraphics[width=0.48\textwidth]{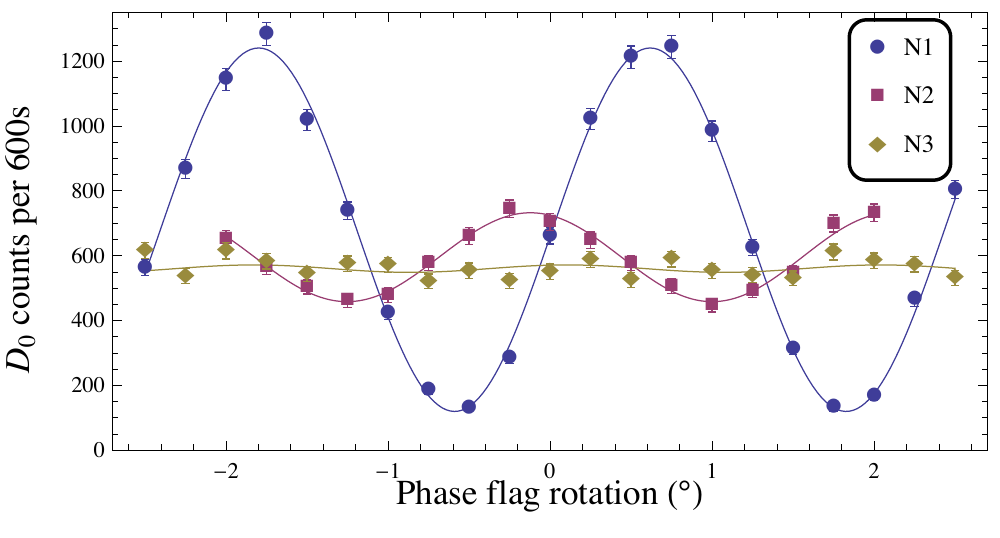}
\caption{(Color online) Measured intensity curves at detector $D_0$ as a function of phase-flag rotation for three NIs. The corresponding path-contrast values are $C_{P1}=(82.5\pm 1.3)\%, C_{P2}=(23\pm1.5)\%, C_{P3}=(2\pm1.7)\%$ for interferometers $N_1, N_2, N_3$ respectively.}
\label{fig:initial-contrast}
\end{center}
\end{figure}

\subsection{Results}
The measured contrast curves in the absence of spin-filtering for the three NIs is shown in Fig. \ref{fig:initial-contrast}, these correspond to contrast values of 
$C_{P1}=(82.5\pm 1.3)\%$, 
$C_{P2}=(23\pm1.5)\%$, 
$C_{P3}=(2\pm1.7)\%$ 
for the interferometers $N_1, N_2$ and $N_3$ respectively. These contrast values correspond to standard deviations of 
$\sigma_1 = 0.62 \pm 0.03 $,
$\sigma_2 = 1.71 \pm 0.04 $,
$\sigma_3 = 2.80 \pm 0.61$ 
respectively in the noise model under consideration.
%

\begin{figure}[htbp]
\begin{center}
\includegraphics[width=0.48\textwidth]{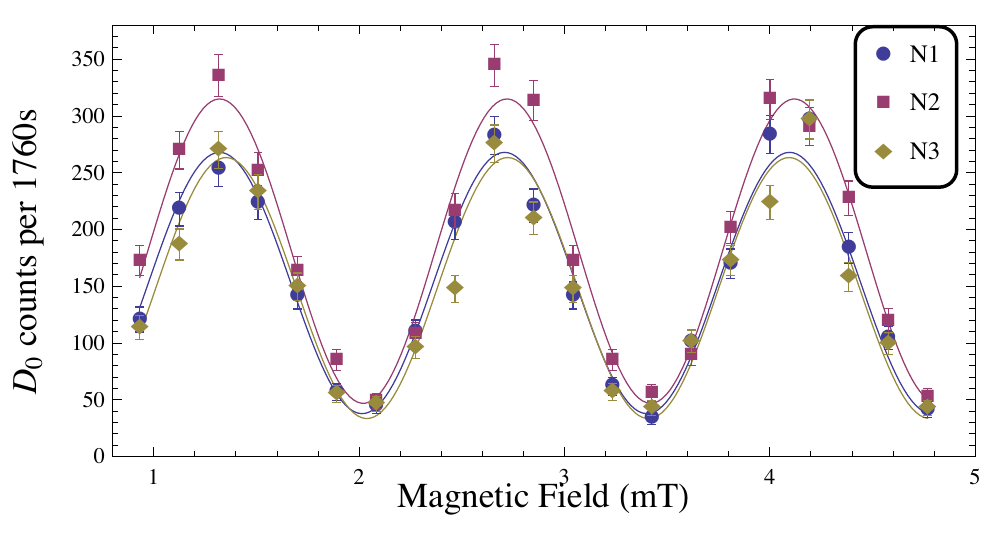}
\caption{(Color online) Measured intensity curves at detector $D_0$ as a function of spin-rotation for three NIs where we have applied a spin-filter on the output beam to select spin-down neutrons with respect to a static magnetic field in the $z$-direction. The corresponding spin-filtered contrast values are $C_{S1(\downarrow z)}=(78.0\pm3)\%, C_{S2(\downarrow z)}=(74.2\pm2.2)\%, C_{S3(\downarrow z)}=(84\pm4)\%$ for interferometers $N_1, N_2, N_3$ respectively.}
\label{fig:spin-contrast}
\end{center}
\end{figure}

After applying the spin-down filter, the spin-filtered contrasts were found to be 
$C_{S1(\downarrow z)}=(78.0\pm3)\%$, 
$C_{S2(\downarrow z)}=(74.2\pm2.2)\%$, 
$C_{S3(\downarrow z)}=(84\pm4)\%$, 
as shown in Fig. \ref{fig:spin-contrast}. Our theoretical model with an initial neutron spin polarization of $(1+\epsilon)/2=93\%$ predicts a spin-contrast of $75.3\%$ for all three interferometers. 

\section{Conclusion}

We have theoretically and experimentally investigated the role of quantum correlations in a simple bipartite quantum system in the presence of noise by using the spin and path degrees of freedom of a polarized neutron beam in a neutron interferometer. If we initially entangle the the path and spin degrees of freedom of a neutron beam by a path dependent spin-rotation, we found that that phase noise acts to reduce the amount of entanglement to zero as the noise strength increases. However a non-zero value of quantum discord $D(A|B)$ for all noise strengths indicates that there are still non-classical correlations between the neutrons spin and path degrees of freedom. The non-zero QD indicates that spin-measurements will have an influence on the quantum state of the neutron path subsystem, however due to the experimental limitations we are only able to perform measurements of the path subsystem in the basis corresponding to the beam paths as implemented by the physical neutron detectors. Restricted to this measurement basis, we are not able to see a noticeable effect for all projective measurements in the strong noise limit. 

In the low noise case our analysis showed that we may think of the spin-path NI as a quantum eraser. In the absence of spin-filtering by rotating the spin state of a neutron in only one path of the interferometer we are labelling the neutrons which take this path and performing a which-way measurement of the neutron's path though the interferometer. This results in a loss of contrast proportional to the entanglement of the path and neutrons. By implementing a post-selected spin measurement in the $x$-direction we may erase this labelling data and restore contrast. This also held true for the spin-contrast, but with the roles of the phase flag and controlled spin rotation angle interchanged. However in the strong noise case, the $X$-filtered path and spin contrast both reduce to zero and so are not observably different from the non-spin-filtered contrast. Thus the effect of $x$-basis spin measurements measurements on the path subsystem state are not directly observable in the NI in the presence of strong dephasing noise. 

In the case of spin-contrast with post-selected spin measurement in the $z$-direction, the contrast remains a function of the spin rotation angle but removes the effect of the phase noise. In the high noise case the expression for spin-contrast when we perform a Z-filter and post-select on the spin down state is a function of spin polarization only. Hence even in the high noise case we are able to experimental observe the effect of spin-filtering on the path subsystem. Our experimental results agree with our theoretical model predicting an increase in spin-filtered contrast over phase contrast for three NIs when spin-filtering has been performed on the Spin-down state in the $z$-direction. The deviations between our measured spin-filtered contrast the value predicted by our theoretical model are consistent with phase variations over the acquisition time due to temperature and humidity fluctuations in the NI environment. We interpreted this non-zero quantum discord as a signature that even in the presence of strong phase noise, the NI still exhibits genuine quantum behaviour.

\acknowledgements
This work was supported by the Canadian Excellence Research Chairs (CERC) program, the Canadian Institute for Advanced Research (CIFAR), and the Natural Sciences and Engineering Research Council of Canada (NSERC) Collaborative Research and Training Experience Program (CREATE). The authors gratefully acknowledge useful discussions with R. Pynn and G.-X. Miao, and use of permalloy films from M. Th. Rekveldt and R. Pynn.

\bibliographystyle{unsrt}

\begin{thebibliography}{}

\bibitem{Nielsen2000}
M.~A. Nielsen and I.~L. Chuang.
\newblock {\em Quantum Computation and Quantum Information}.
\newblock Cambridge University Press, 2000.

\bibitem{Horodecki2009}
R. Horodecki, P. Horodecki, M. Horodecki, and K. Horodecki.
\newblock Quantum entanglement.
\newblock {\em Rev. Mod. Phys.}, 81, 2009.

\bibitem{Lang2011}
M.~D. Lang, C.~M. Caves, and A.~Shaji.
\newblock Entropic measures of non-classical correlations.
\newblock {\em International Journal of Quantum Information}, 09(07n08), 2011.

\bibitem{Celeri2011}
L.~C. C{\'e}leri, J.~Maziero, and R.~M. Serra.
\newblock Theoretical and experimental aspects of quantum discord and related
  measures.
\newblock {\em International Journal of Quantum Information}, 09(07n08), 2011.

\bibitem{Modi2011}
K.~Modi, A.~Brodutch, H.~Cable, T.~Paterek, and V.~Vedral.
\newblock The classical-quantum boundary for correlations: Discord and related
  measures.
\newblock {\em Rev. Mod. Phys.}, 84, 2012.

\bibitem{Ollivier2002}
H.~Ollivier and W.~H. Zurek.
\newblock {Quantum discord: A measure of the quantumness of correlations}.
\newblock {\em Phys. Rev. Lett.}, {88}, {2002}.

\bibitem{Henderson2001}
L.~Henderson and V.~Vedral.
\newblock Classical, quantum and total correlations.
\newblock {\em J. Phys. A}, 34, 2001.

\bibitem{Knill1998}
E.~Knill and R.~Laflamme.
\newblock Power of one bit of quantum information.
\newblock {\em Phys. Rev. Lett.}, 81, 1998.

\bibitem{Datta2008}
A.~Datta, A.~Shaji, and C.~M. Caves.
\newblock Quantum discord and the power of one qubit.
\newblock {\em Phys. Rev. Lett.}, 100, 2008.

\bibitem{Colella1975}
R.~Colella, A.~W. Overhauser, and S.~A. Werner.
\newblock Observation of gravitationally induced quantum interference.
\newblock {\em Phys. Rev. Lett.}, 34, 1975.

\bibitem{Rauch1975}
H.~Rauch, A.~Zeilinger, G.~Badurek, A.~Wilfing, W.~Bauspiess, and U.~Bonse.
\newblock Verification of coherent spinor rotation of fermions.
\newblock {\em Physics Letters A}, 54, 1975.

\bibitem{Summhammer1983}
J.~Summhammer, G.~Badurek, H.~Rauch, U.~Kischko, and A.~Zeilinger.
\newblock Direct observation of fermion spin superposition by neutron
  interferometry.
\newblock {\em Phys. Rev. A}, 27, 1983.

\bibitem{Cimmino1989}
A.~Cimmino, G.~I. Opat, A.~G. Klein, H.~Kaiser, S.~A. Werner, M.~Arif, and
  R.~Clothier.
\newblock Observation of the topological aharonov-casher phase shift by neutron
  interferometry.
\newblock {\em Phys. Rev. Lett.}, 63, 1989.

\bibitem{Hasegawa2003}
Y.~Hasegawa, R.~Loidl, G.~Badurek, M.~Baron, and H.~Rauch.
\newblock Violation of a bell-like inequality in single-neutron interferometry.
\newblock {\em Nature}, 425, 2003.

\bibitem{Hasegawa2007}
Y. Hasegawa, R. Loidl, G. Badurek, S. Filipp, J. Klepp, and
  H. Rauch.
\newblock Evidence for entanglement and full tomographic analysis of bell
  states in a single-neutron system.
\newblock {\em Phys. Rev. A}, 76:052108, Nov 2007.

\bibitem{Bartosik2009}
H.~Bartosik, J.~Klepp, C.~Schmitzer, S.~Sponar, A.~Cabello, H.~Rauch, and
  Y.~Hasegawa.
\newblock Experimental test of quantum contextuality in neutron interferometry.
\newblock {\em Phys. Rev. Lett.}, 103, 2009.

\bibitem{Pushin2011}
D.~A. Pushin, M.~G. Huber, M.~Arif, and D.~G. Cory.
\newblock Experimental realization of decoherence-free subspace in neutron
  interferometry.
\newblock {\em Phys. Rev. Lett.}, 107, 2011.

\bibitem{Galve2011}
F.~Galve, G.~L. Giorgi, and R.~Zambrini.
\newblock Orthogonal measurements are almost sufficient for quantum discord of
  two qubits.
\newblock {\em EPL}, 96, 2011.

\bibitem{Wootters1998}
W.~K. Wootters.
\newblock {Entanglement of Formation of an Arbitrary State of Two Qubits}.
\newblock {\em Physical Review Letters}, 80, 1998.

\bibitem{Zhou2012}
D.~Zhou, G-W. Chern, J.~Fei, and R.~Joynt.
\newblock Topology of entanglement evolution of two qubits.
\newblock {\em International Journal of Modern Physics B}, 26, 2012.

\bibitem{Yu2009}
T. Yu and J.~H. Eberly.
\newblock Sudden death of entanglement.
\newblock {\em Science}, 323(5914), 2009.

\bibitem{Werlang2009}
T.~Werlang, S.~Souza, F.~F. Fanchini, and C.~J. Villas~Boas.
\newblock Robustness of quantum discord to sudden death.
\newblock {\em Phys. Rev. A}, 80, 2009.

\bibitem{Ferraro2010}
A.~Ferraro, L.~Aolita, D.~Cavalcanti, F.~M. Cucchietti, and A.~Ac{\'\i}n.
\newblock Almost all quantum states have nonclassical correlations.
\newblock {\em Phys. Rev. A}, 81, 2010.

\bibitem{Fanchini2010}
F.~F. Fanchini, L.~K. Castelano, and A.~O. Caldeira.
\newblock Entanglement versus quantum discord in two coupled double quantum
  dots.
\newblock {\em New Journal of Physics}, 12(7), 2010.

\bibitem{ZhangYJ2011}
Y-J Zhang, X-B Zou, Y-J Xia, and G-C Guo.
\newblock Quantum discord dynamics in the presence of initial system--cavity
  correlations.
\newblock {\em Journal of Physics B: Atomic, Molecular and Optical Physics},
  44(3), 2011.

\bibitem{Sears1989}
V.~F. Sears.
\newblock {\em Neutron Optics}.
\newblock Oxford University Press, 1989.

\bibitem{Scully1982}
M.~O. Scully, and K.~Dr{\"u}hl.
\newblock Quantum eraser: A proposed photon correlation experiment concerning observation and "delayed choice" in quantum mechanics.
\newblock {\em Phys. Rev. A}, 25, 1982.

\bibitem{Aharonov2013}
Y. Aharonov, S. Popescu, D. Rohrlich, P. Skrzypczyk
\newblock {Quantum Cheshire Cats}.
\newblock {\em New Journal of Physics}, 15, 113015, 2013.

\bibitem{Denkmayr2013}
T. Denkmayr, H. Geppert, S. Sponar, H. Lemmel, A. Matzkin, J. Tollaksen, Y. Hasegawa.
\newblock {Observation of a quantum Cheshire Cat in a matter wave interferometer experiment}.
\newblock {\em arXiv e-prints}, arXiv:1312.3775 [quant-ph], 2013.

\bibitem{NistWeb}
{Neutron Interferometry and Optics Facility}.
\newblock \url{http://physics.nist.gov/majresfac/interfer/text.html}.

\bibitem{Pushin2008}
D.~A. Pushin, M.~Arif, M.~Huber, and D.~G. Cory.
\newblock {Measurements of the Vertical Coherence Length in Neutron
  Interferometry}.
\newblock {\em Physical Review Letters}, 100, 2008.

\bibitem{Abutaleb2012}
M.~O. Abutaleb, D.~A. Pushin, M.~G. Huber, C.~F. Majkrzak, M.~Arif, and D.~G.
  Cory.
\newblock Design of remnant magnetization fecov films as compact, heatless
  neutron spin rotators.
\newblock {\em Applied Physics Letters}, 101, 2012.

\bibitem{PynnFootnote}
{On loan from R. Pynn}

\bibitem{Pynn2005}
Roger Pynn.
\newblock Broadband spin flippers constructed from thin magnetic films.
\newblock {\em Physica B: Condensed Matter}, 356, 2005.

\bibitem{Pushin2007}
D.~A. Pushin, D.~G. Cory, M.~Arif, D.~L. Jacobson, and M.~G. Huber.
\newblock {Reciprocal space approaches to neutron imaging}.
\newblock {\em Applied Physics Letters}, 90, 2007.


\end{thebibliography}

\end{document}